\documentclass[%
 reprint,
superscriptaddress,
 amsmath,amssymb,
 aps,
]{revtex4-2}

\usepackage{graphicx}
\usepackage{dcolumn}
\usepackage{bm}
\usepackage{tikz}
\usepackage{xcolor}
\usepackage{amsmath}
\usepackage{csquotes}

\newcommand{\p}{\partial}
\newcommand{\s}{\sigma}
\newcommand{\Mc}{\mathcal{M}}
\newcommand{\Bc}{\mathcal{B}}
\newcommand{\ri}{\mathrm{i}}
\newcommand{\dd}{\mathrm{d}}
\newcommand{\Qh}{\widehat{Q}}
\newcommand{\Hh}{\widehat{H}}
\newcommand{\Tr}{\text{Tr}}
\newcommand{\slc}{\mathfrak{sl}(2,\mathbb{C})}
\newcommand{\Lc}{\mathcal{L}}

\definecolor{myred}{rgb}{0.75, 0.25, 0}
\definecolor{mygreen}{rgb}{0.25, 0.75, 0}
\definecolor{myblue}{rgb}{0, 0.25, 0.75}
\definecolor{mypurple}{rgb}{0.6, 0.2, 0.8}

\let\originalleft\left
\let\originalright\right
\renewcommand{\left}{\mathopen{}\mathclose\bgroup\originalleft}
\renewcommand{\right}{\aftergroup\egroup\originalright}
\newcommand{\br}[1]{\left( #1 \right)}

\newcommand{\re}{\mathrm{e}}

\begin{document}

\title{Ricci flow of Integrable Sigma Models, Hitchin Systems\\ and Quantum Seiberg--Witten theory
}
\thanks{The sausage reheated!}%

\author{G. Bonelli}
\email{bonelli@sissa.it}
\affiliation{%
  SISSA, Via Bonomea 265, IGAP, via Beirut 2, INFN, Sezione di Trieste, Trieste, Italy}

\author{S. Lacroix}
\email{sylvain.lacroix@sorbonne-universite.fr}
\affiliation{%
  Sorbonne Université, CNRS, Laboratoire de Physique Théorique et Hautes Énergies, \\
  LPTHE, F-75005 Paris, France}

\author{T. Pedroni}
\email{tommaso.pedroni@unige.ch}
\affiliation{%
  SISSA, Via Bonomea 265, IGAP, via Beirut 2, INFN, Sezione di Trieste, Trieste, Italy}

\affiliation{%
  Section de Mathématiques, Université de Genève,
  1211 Genève 4, Switzerland}

\author{A. Tanzini}
\email{tanzini@sissa.it}
\affiliation{%
  SISSA, Via Bonomea 265, IGAP, via Beirut 2, INFN, Sezione di Trieste, Trieste, Italy}

\date{\today}

\begin{abstract}
\noindent We construct a family of integrable 2d sigma models through 4d Chern--Simons theory and show that their classical point-particle reduction corresponds to a Hitchin system. The integrable Hitchin quantisation then provides a natural candidate Hamiltonian for the one-loop reduced dynamics of the sigma models, which coincides with the Perelman operator in all cases examined. This allows us to study the corresponding spectral problems using 4d $\mathcal{N}=2$ gauge theory and 2d conformal field theory methods. For the sausage model, we derive exact quantisation conditions and compute exponentially small corrections essential for the agreement with numerical finite-volume results.

\end{abstract}

\maketitle

\section{Introduction}

The relation between gauge theories and integrable systems has a long and successful history, with Seiberg--Witten theory as one of its most important manifestations~\cite{Seiberg_1994,Seiberg_1994b}. The low-energy dynamics of a four-dimensional $\mathcal N=2$ gauge theory is encoded in an algebraic curve and a meromorphic differential, naturally identified with the spectral curve of a classical integrable system~\cite{Gorsky:1995zq,Donagi_1996}. For class $\mathcal S$ theories this is the Hitchin system on a punctured Riemann surface~\cite{Gaiotto:2009we,Bonelli:2009zp}. Equivariant localization relates this geometry to quantum integrability: in the Nekrasov--Shatashvili (NS) limit, the Nekrasov partition function of the theory determines the spectrum of the quantized system~\cite{Nekrasov:2009rc,Nekrasov_2011}. Through the AGT correspondence~\cite{Alday:2009aq}, this also gives a two-dimensional conformal field theory (2d CFT) description of quantum Hitchin Hamiltonians.
A more recent instance of this gauge-theoretic approach to integrability is the construction of two-dimensional integrable field theories from four-dimensional Chern--Simons (4d CS) theory~\cite{Costello:2019tri,Lacroix:2020flf}. 
There, a meromorphic one-form on a Riemann surface, together with suitable defects, determines a classical integrable model in two dimensions. This construction is closely related~\cite{Vicedo:2019dej} to affine Gaudin models, where the pole structure encodes the algebraic data of the Lax connection~\cite{Feigin:2007mr,Vicedo:2017cge,Delduc:2019bcl}. In particular, it provides a systematic way to engineer integrable non-linear sigma models (iNLSMs), whose metric and $B$-field satisfy the one-loop renormalization group (RG) equations~\cite{Delduc:2020vxy,Derryberry:2021rne,Lacroix:2025ias}.

In this Letter we explore new correspondences between these approaches, relating the 1-loop RG and spectrum of a class of iNLSMs to the quantum Seiberg--Witten/Hitchin system. We show that, upon a generalized point-particle reduction (sometimes also called minisuperspace approximation~\cite{Fateev:1992tk}), the classical zero-mode dynamics of these NLSMs can be recast in terms of Hitchin Hamiltonians on the punctured sphere associated with the underlying affine Gaudin model \footnote{See also~\cite{Schepers:2020ehn} and~\cite{Bykov:2025lcb} for similar observations in the cases of the 3d sausage and the flag manifold $U(3)/U(1)^3$.}. This observation gives a natural way to address the quantisation of the reduced theory, which would otherwise suffer from the usual ambiguities of quantum mechanics on a curved target space. We propose to resolve these ambiguities by using the Hitchin quantisation \cite{Gaudin1976,1885368} of the classical Hamiltonian, which selects a distinguished quantum operator preserving integrability, that we expect to encode the 1-loop spectrum of the NLSM particle sector.

We provide evidence supporting this proposal. First, in all the cases that we studied, \textit{i.e.} certain Hitchin systems with up to 6 punctures, the resulting quantum Hamiltonian coincides with the Perelman operator \cite{perelman2002entropyformularicciflow,Oliynyk:2005ak} associated with the Ricci-flow of the target space. This gives a precise link between quantum Hitchin integrability and the one-loop RG-flow of the iNLSMs, with the lowest eigenvalue of this operator being related to the $c$-function~\cite{Oliynyk:2005ak,Tseytlin:2006ak}.
Second, this identification makes the spectral problem of the Hamiltonian accessible to analytic tools from supersymmetric localization and the AGT correspondence. For the sausage model~\cite{Fateev:1992tk}, we use exact quantisation conditions obtained from connection formulae derived by two-dimensional CFT methods, with conformal blocks expressed exactly in terms of Nekrasov functions~\cite{Bonelli:2022ten}. This allows us to compute exponentially small corrections to the eigenvalues, see Eq.~\eqref{eq:double-expansion-alpha}, which are essential for precise agreement with numerical results obtained from the non-linear integral equations governing the sausage model spectrum~\cite{Bazhanov:2017nzh}. This matching provides concrete evidence for the proposed Hitchin quantisation of the point-particle sector of these iNLSMs and demonstrates the effectiveness of localization-based methods in the study of Perelman operators. A more systematic treatment of the general framework and of the technical details will be presented elsewhere \cite{Hitchin}. 

Several directions deserve further investigation. The appearance of Hitchin systems in both the class $\mathcal S$ description of four-dimensional $\mathcal N=2$ theories and in the point-particle quantisation of the iNLSMs considered here suggests that the correspondence may have a common higher-dimensional origin, possibly arising from string/M-theory constructions in different compactification limits.
A second important question concerns the spectral problem itself. Although our construction identifies the Perelman operator as a quantum Hitchin Hamiltonian, the choice of domain, or equivalently of the appropriate real slice and boundary conditions for the eigenfunctions, is not fixed by the operator alone. The Hitchin-system perspective may provide a natural prescription for this choice, potentially through its relation to opers and the analytic Langlands correspondence \cite{Teschner:2017djr,Gaiotto:2021tsq,Etingof:2023drx}. 
Another direction is suggested by the classical circle reduction of 4d CS theory, whose bulk action is a holomorphic-topological 3d BF theory, and could provide a gauge-theoretic interpretation of the quantum point-particle approximation by quantising the setup of~\cite{Vicedo:2022mrm}. One may also try to find a supersymmetric extension of the 4d CS theory compatible with the pole and defect conditions engineering the iNLSM, together with an equivariant supercharge whose square generates spatial translations along the worldsheet cylinder, modulo gauge and R-symmetry transformations. If a suitable localization construction exists, it could provide a path-integral description of the circle reduction and a relation to the point-particle sector of the supersymmetric model. Finally, it would be important to understand to what extent the constructions considered in this Letter can be extended to include higher-loop corrections to the iNLSMs studied here. It is worth noting that an equivalent of the Perelman operator for the higher-loop RG-flow has been studied in~\cite{Tseytlin:2006ak}. It would be interesting to investigate whether the spectrum of this operator is related to the higher-loop spectrum of the iNLSMs in the point-particle sector and to their higher-loop $c$-function. Note that the all-loop RG-flow of planar iNLSMs has also been studied recently from 4d CS and holography~\cite{Costello:2026gve}.
We hope to return to these questions in future work.

\section{Integrable NLSMs and 1-loop point-particle reduction}
\label{sec:iNLSM}

\paragraph{iNLSMs.} Consider a NLSM with action
\begin{equation}\label{eq:NLSM}
S[X] = \frac{1}{2}\iint_\Sigma\text{d} \tau\,\text{d}\sigma\, \bigl( G_{\mu\nu}(X) + B_{\mu\nu}(X) \bigr) \p_+ X^\mu\,\p_- X^\nu\,.    
\end{equation}
Here $X^\mu(\tau,\sigma)$ ($\mu=1,\dots,d$) are the coordinate fields, $(\tau,\sigma)$ parameterise the 2d cylindrical worldsheet $\Sigma$ with $\sigma\sim\sigma+L$ periodic, $\p_\pm = \p_\tau\pm\p_\s$ are light-cone derivatives while $G_{\mu\nu}$ and $B_{\mu\nu}$ are the (pseudo-Riemannian) metric and B-field of the target space $\mathcal{M}$. We will focus on a specific class of iNLSMs admitting a Lax connection $\mathcal{L}(z) = \mathcal{L}_\tau(z) \text{d}\tau+\mathcal{L}_\sigma(z)\text{d}\sigma$. The detailed construction and analysis of these iNLSMs will be spelled out elsewhere~\cite{Hitchin}: we summarise their main features here and gather some useful facts in Appendix \ref{app:NLSM}. They are built from the unifying frameworks of $\mathrm{SL}(2,\mathbb{C})$ affine Gaudin models~\cite{Feigin:2007mr,Vicedo:2017cge,Delduc:2019bcl} / 4d CS theory~\cite{Costello:2019tri,Lacroix:2020flf}, which automatically ensures their integrability \footnote{For simplicity, we focus here on integrable models associated with the Lie group $\mathrm{SL}(2,\mathbb{C})$ and a rational spectral parameter in $\mathbb{CP}^1$, but we expect that similar results can be obtained for higher-rank groups and possibly higher-genus curves. Moreover, we focus only on a subclass of all possible iNLSMs obtained through these frameworks.}. The spectral parameter $z\in\mathbb{CP}^1$ plays a key role in this construction: in particular, the models depend on the choice of a meromorphic \enquote{twist 1-form} $\omega=\omega(z)\text{d} z$ on $\mathbb{CP}^1$ with simple poles and zeroes, which are the main defining parameters of the theory. Moreover, these iNLSMs are 1-loop renormalisable. In other words, the target space geometry $(\mathcal{M},G,B)$ is a solution of the torsion-full Ricci-flow, with the RG-flow being written compactly as a running of $\omega$~\cite{Delduc:2020vxy,Derryberry:2021rne,Lacroix:2025ias,Costello:2026gve}. The simplest example among this class of iNLSMs is the sausage model: see Appendix \ref{app:NLSM} and Section \ref{sec:Sausage} for more details. The iNLSMs that we consider admit a large number of abelian isometries, namely we can choose a chart in which $G_{\mu\nu}$ and $B_{\mu\nu}$ depend only on the first $n$ coordinates, with $n<d=\dim(\Mc)$. Geometrically, we have a bundle $\Mc\to\Bc$ (locally), with the base $\Bc$ described by the coordinates $(X^1,\dots,X^n)$ and the fibre by $(X^{n+1},\dots,X^d)$, such that the metric and B-field are constant on each fibre. The dimension of the base $n=\text{dim}(\Bc)$ is equal to the number of poles of $\omega$ minus 3. We will reserve labels $r,s\in\lbrace 1,\dots,n \rbrace$ for base indices and $a,b\in\lbrace n+1,\dots,d\rbrace$ for fibre ones. One can consider different topologies for the fibres: either compact, when all the $X^a$'s are periodic, or non-compact. For instance, the sausage is described by a non-trivial $S^1$-fibration over the interval, the fibre shrinking at the end points, and is therefore homeomorphic to a two-sphere.\vspace{5pt}

\paragraph{Classical particle.} We consider a point-particle limit (also called minisuperspace approximation~\cite{Fateev:1992tk}) of the iNLSM by focusing on field configurations $X^\mu(\tau,\sigma) = x^\mu(\tau)$. This describes a subsector of the iNLSM solutions in terms of a mechanical model with variables $x^\mu(\tau)$ and Lagrangian $\frac{L}{2} G_{\mu\nu}(x) \dot{x}^\mu \dot{x}^\nu$. The classical Hamiltonian is then $\frac{1}{L}H^{\Mc}$, with $H^{\Mc}=\frac{1}{2}G^{\mu\nu}(x)p_\mu p_\nu$. By construction, the fibre coordinates $x^a$ are cyclic and their momenta $p_a$ are conserved. We can thus restrict to solutions with $p_a$ fixed to given constants $\pi_a$. This results in a system with canonical coordinates $(x^r,p_r)_{r=1,\dots,n}$ and Hamiltonian
\begin{equation}
    H = \frac{1}{2} G^{rs}(x) p_r p_s + \pi_aG^{ra}(x)p_r + \frac{1}{2} \pi_a \pi_b G^{ab}(x) \,,
\end{equation}
with $\pi_a$ treated as parameters.

Our first claim is that this model can be identified with (a real-slice of) a $\mathrm{SL}(2,\mathbb{C})$ Hitchin system on a $(n+3)$-punctured sphere. This is an integrable model~\cite{Hitchin:1987mz} with canonical coordinates $(x^r,p_r)$, depending on certain geometric parameters: punctures on $\mathbb{CP}^1$ and their attached parabolic weights. Its Poisson-commuting Hamiltonians are extracted as coefficients of a meromorphic quadratic differential $Q(z)\dd z^2$ with poles at the punctures, satisfying $\lbrace Q(z), Q(w) \rbrace=0$ with respect to the canonical bracket. In the relation with the above iNLSM, the punctures are identified with the poles of $\omega$, while the parabolic weights are expressed in terms of the fibre parameters $\pi_a$, in a way which we will spell out in~\cite{Hitchin}. We then find that the reduced particle Hamiltonian can be written in terms of the Hitchin differential as
\begin{equation}\label{eq:HHitchin}
    H = \sum_{i=1}^{n+1} \pm \mathop{\mathrm{res}}_{z=\zeta_i} \frac{Q(z)}{\omega(z)}\text{d} z\,,
\end{equation}
where $\zeta_i$ are the zeroes of $\omega(z)\dd z$ and $\pm$ is a suitable choice of sign attached to each $\zeta_i$, arising from the 4d CS construction --- see Appendix~\ref{app:NLSM}. The above equation then extracts $H$ as a specific combination of the Hitchin Hamiltonians, with coefficients that also depend on the zeroes of $\omega$. The latter are parameters of the original iNLSM, but not of the Hitchin system. The classical point-particle reduction thus produces an integrable mechanical system, as $H$ Poisson-commutes with all Hitchin Hamiltonians. Moreover, the combination $L\,\omega(z)\mathcal{L}_\sigma(z)\text{d} z$ built from the iNLSM Lax connection reduces to the so-called Higgs field of the Hitchin system, while the zero curvature equation of $\mathcal{L}_{\tau/\sigma}(z)$ becomes its isospectral evolution. We refer to Appendix~\ref{app:NLSM} for an illustration of that relation in the sausage model. 

The above relation between 2d iNLSMs and Hitchin systems, and in particular the identity~\eqref{eq:HHitchin}, is established~\cite{Hitchin} using the Hamiltonian formalism, by reducing affine Gaudin models to finite Gaudin models. Equivalently, we expect that it can be understood from the gauge theory origin of the iNLSMs. Namely, as outlined in Appendix~\ref{app:NLSM}, it should arise through the reduction of the 4d CS theory to a 3d topological-holomorphic BF theory. Indeed, the former localises to iNLSMs while the latter is known to produce Hitchin systems~\cite{Vicedo:2022mrm}.\vspace{5pt}

\paragraph{One-loop point-particle approximation.} We now turn to the quantum theory. Our main focus will be the finite-volume spectrum, \textit{i.e.} the energy of the quantum iNLSM states as a function of the cylinder size $L$. A full rigorous and non-perturbative answer to that question is clearly out of reach and we thus simplify the problem by working in the 1-loop point-particle approximation~\cite{Fateev:1992tk}. The main expectation is that the eigenfunctions $\Psi_\ell^{\Mc}(x)$ of the quantisation of the particle Hamiltonian $H^{\Mc}$ correspond to QFT eigenstates in the zero worldsheet momentum subsector of the full iNLSM Hilbert space, with their energy spectrum given, at 1-loop, by
\begin{equation}\label{eq:Energy}
    E_\ell(L) = -\frac{\pi d}{6 L} + \frac{\varepsilon_\ell(L)}{L}  \,.
\end{equation}
The first term is the standard Casimir energy for $d=\dim(\Mc)$ free bosons on a cylinder, while $\varepsilon_\ell(L)$ is the eigenvalue of the quantum particle Hamiltonian, which depends on the target geometry and thus captures the interactions in the NLSM. The $L$-dependence in the eigenvalue $\varepsilon_\ell(L)$ is introduced by letting the background $(\mathcal{M},G,B)$ run according to the 1-loop RG-flow, with the cylinder size setting the physical energy scale $L^{-1}$. This is a remnant of the field-theoretical nature of the NLSM in its point-particle sector.

A non-trivial question here is the choice of quantised particle Hamiltonian. Recall that, before eliminating the fibre coordinates, the classical Hamiltonian is $H^{\Mc}=\frac{1}{2}G^{\mu\nu}(x)p_\mu p_\nu$. This is famously sensitive to ordering ambiguities at the quantum level: indeed, $p_\mu$ gets promoted to the operator $-\ri\p_{x^\mu}$, which does not commute with $G^{\mu\nu}(x)$. A standard covariant choice of ordering for the quantum Hamiltonian is $-\frac{1}{2}\Delta$, where $\Delta$ is the Laplace-Beltrami operator
\begin{equation}
    \Delta = \frac{1}{\sqrt{|G|}} \p_\mu\bigl( \sqrt{|G|}\, G^{\mu\nu}\p_\nu \bigr)\,.
\end{equation}
Other natural quantisations lead to corrections by a potential proportional to the scalar curvature. More generally, any operator of the form
\begin{equation}
    \Hh^{\Mc}_V =-\frac{1}{2}\Delta + V(x)\,,
\end{equation}
with $V(x)$ a scalar potential on $\Mc$, is a quantisation of the same classical Hamiltonian $H^{\Mc}$. The question is therefore the following: which quantisation $\Hh^{\Mc}_V$ is the correct one to describe the NLSM spectrum in~\eqref{eq:Energy} (at 1-loop)? In the present case, the classical particle model coincides with an integrable Hitchin system. We will thus take this as a guide to find $\Hh^{\Mc}_V$. Indeed, it is well-known how to quantise the Hitchin system in a way which preserves integrability~\cite{Gaudin1976,1885368}. In this construction, the quadratic differential $Q(z)$ gets promoted to a quantum operator $\Qh(z)$, satisfying $\bigl[\Qh(z),\Qh(w)\bigr]=0$.

Note that $\Hh^{\Mc}_V$ and $\Qh(z)$ are differential operators on the full target space $\Mc$ and on the base $\Bc$, respectively. To be able to compare them, we then first need to perform the reduction of $\Hh^{\Mc}_V$ onto the base. At the classical level, this was done by setting the fibre momenta $p_a$ to constants $\pi_a$. The quantum equivalent of this procedure is to restrict to a wavefunction
\begin{equation}\label{eq:QuantRed}
    \Psi^{\Mc}_\ell(x^1,\dots,x^d) = \re^{\ri \pi_a x^a} \Psi_\ell(x^1,\dots,x^n)\,,
\end{equation}
chosen to be an eigenvector of $-\ri\p_{x^a}$ with eigenvalues $\pi_a$, which are discrete for compact fibres and continuous for non-compact ones. 
Since the geometry does not depend on the fibre coordinates $x^a$ (and assuming that $V$ does not either), the eigenvalue equation $\Hh^{\Mc}_V \Psi^{\Mc}_\ell = \varepsilon_\ell\,\Psi^{\Mc}_\ell$ translates to $\Hh_V \Psi_\ell = \varepsilon_\ell\Psi_\ell$, with $\Hh_V$ a differential operator on the base $\Bc$, depending on the parameters $\pi_a$.  We can now state the main conjecture of this section.\vspace{5pt}

\noindent \textbf{Conjecture:} The quantised particle Hamiltonian is related to the quantised Hitchin operator by
\begin{equation}\label{eq:ConjQ}
    |G|^{-1/4}\,\Hh_V\, |G|^{1/4} = \sum_{i=1}^{n+1} \pm \text{res}_{z=\zeta_i} \frac{\Qh(z)}{\omega(z)}\text{d} z + c_0\,,
\end{equation}
if we choose the scalar potential
\begin{equation}\label{eq:IntPot}
    V(x) = \frac{1}{8} R(x) - \frac{1}{96} H_{\mu\nu\rho}(x)\,H^{\mu\nu\rho}(x)\,,
\end{equation}
where $R(x)$ is the Ricci scalar, $H=\dd B$ is the torsion of the NLSM and $c_0$ is a model-dependent constant.\vspace{5pt}

We propose to use this specific choice of $\Hh_V$ to describe the 1-loop NLSM spectrum in \eqref{eq:Energy}. We note that the same operator was used in~\cite{Fateev:1992tk,Fateev:2018yos} for the sausage model and higher-rank generalisations. A few remarks are in order. The conjecture in~\eqref{eq:ConjQ}--\eqref{eq:IntPot} is the quantum equivalent of~\eqref{eq:HHitchin} and we have checked it on the example of the sausage model (see next section) as well as other cases with up to 6 punctures (see~\cite{Hitchin}). We note that the conjugation by $|G|^{1/4}$ does not affect the spectrum and the integrability properties of the Hamiltonian.
Here, the choice of potential in~\eqref{eq:IntPot} is guided by the identification with the Hitchin operator and thus by quantum integrability. It is worth noting that it involves the NLSM's torsion (which was necessary for its integrability and renormalisability), even though the classical point-particle is insensitive to this object. It is therefore another remnant of the field-theoretical nature of the full quantum NLSM. But the most remarkable observation is the following: the specific potential in~\eqref{eq:IntPot} has a very important geometric interpretation. Namely, the lowest eigenvalue $\varepsilon_0$ of the associated operator $\Hh^{\Mc}_V$ is the Perelman entropy~\cite{perelman2002entropyformularicciflow,Oliynyk:2005ak} of the background $(\Mc,G,B)$, which is a well-known functional whose gradient flow is the torsion-full Ricci flow, \textit{i.e.} the 1-loop RG-flow of the NLSM. This suggests that the Perelman operator might be a natural prescription for the 1-loop Hamiltonian of the particle sector of a renormalisable NLSM.
In particular, this prescription seems compatible with integrability and the connection to Hitchin systems then provides very powerful tools to study the spectrum of this operator (including relations to $\mathcal{N}=2$ gauge theories~\cite{Nekrasov:2009rc,Nekrasov_2011} and the Langlands correspondence~\cite{1885368,Teschner:2017djr,Gaiotto:2021tsq,Etingof:2023drx}). We will see an explicit illustration of this in Section \ref{sec:Sausage}, with the example of the sausage model (see also~\cite{Schepers:2020ehn} for a complementary approach using the WKB method in the 3d sausage model). In particular, this case provides some numerical evidence supporting the energy formula in~\eqref{eq:Energy}, with the Hamiltonian in Eqs.~\eqref{eq:ConjQ}--\eqref{eq:IntPot}.\vspace{5pt}

\paragraph{Adding \enquote{winding/twist}.} We can generalise the point-particle reduction by introducing a mild $\s$-dependence in the classical ansatz:
\begin{equation}
\label{eq:generalized_minisuperspace}
    X^r(\tau,\sigma) = x^r(\tau)\,, \qquad X^a(\tau,\sigma)=x^a(\tau)+\mu^a\sigma\,,
\end{equation}
for $r\in\lbrace 1,\dots,n \rbrace$ and $a\in\lbrace n+1,\dots,d\rbrace$, with $\mu^a$ being constant parameters. The standard point-particle ansatz corresponds to $\mu^a=0$. A non-zero $\mu^a$ encodes winding, twist or quasi-periodicity along the fibre coordinate $X^a$, depending on the global geometry of the bundle $\Mc\to\Bc$ \footnote{If the fibre coordinate $X^a\sim X^a+\Lambda^a$ is compact with period $\Lambda^a$ and the NLSM field is taken to be periodic in $\s\sim\s+L$, then $\mu^a=w^a \Lambda^a/L$ where $w^a\in\mathbb{Z}$ is the standard integer winding number. More generally, allowing the field to satisfy a quasi-periodic boundary condition results in $\mu^a=(u^a+w^a)\Lambda^a/L$ with a fixed number $u^a\in[0,1)$ and $w^a\in\mathbb{Z}$ quantised. If the fibre coordinate is not compact, $\mu^a$ is not quantised.}. The key observation is that the NLSM action, equations of motion and Lax connection depend on the derivatives $\p_\pm X^a$ of the fibre coordinates but not on the $X^a$ themselves. Therefore, the $\s$-dependence drops out and the above ansatz is consistent. We then obtain a deformed mechanical model, depending on the NLSM background $(\Mc,G,B)$ and the parameters $(\pi_a,\mu^a)$. We note that the exchange of $\pi_a$ and $\mu^a$ essentially encodes T-duality along the fibre.

Most of our previous claims/conjectures extend to this generalised case. The classical model can be identified with a Hitchin system as in~\eqref{eq:HHitchin}. The only difference is in the choice of parabolic weights at the punctures, which now also depend on the parameters $\mu^a$ in addition to $\pi_a$ (see~\cite{Hitchin} for explicit formulas as well as the next section and Appendix~\ref{app:NLSM} for the example of the sausage). This Hitchin system can still be quantised while preserving integrability and this results in a natural proposal for the 1-loop mechanical Hamiltonian governing the winding/twisted sector of the NLSM Hilbert space. 

\section{The Sausage Model}
\label{sec:Sausage}

In this section, we illustrate the general construction above using the sausage model, for which the relevant spectral problem can be studied explicitly. Further details will be presented in~\cite{Hitchin}. The sausage model~\cite{Fateev:1992tk} is an iNLSM with two-dimensional target-space metric
\begin{equation}
\label{eq:sausage_metric_rational}
    \dd s^2 = \frac{1}{2\nu(\mathfrak{t}-X)} \left( \frac{\dd X^2}{X(1-X)} + 4X(1-X)\dd Y^2 \right)\,, 
\end{equation}
where $X\in(0,1)$ and $Y\sim Y+2\pi$ --- see Appendix~\ref{app:NLSM} for its 4d CS origin and the explicit formulae of the twist 1-form, Lax connection, Higgs field and classical Hamiltonian of the associated Hitchin system. Here $\nu\ll1$ is the semiclassical loop-counting parameter, while $\mathfrak{t}=\mathfrak{t}(\eta)$ is the running modulus of the sausage defined in~\eqref{eq:rational_variables}, increasing from $\mathfrak{t}=1^{+}$ in the UV to $\mathfrak{t}\to +\infty$ in the IR. On the worldsheet cylinder with $\sigma\sim\sigma+L$, we consider the generalized point-particle ansatz
\begin{equation}
\label{eq:minisuperspace_ansatz}
X\br{\tau,\sigma}=x(\tau), \qquad Y(\tau,\sigma)=y(\tau)+\frac{\nu \omega_W}{2}\frac{\sigma}{L},
\end{equation}
where $\omega_W$ is the winding/twist parameter. In the 1-loop point-particle approximation, the corresponding finite-volume eigenvalues $\varepsilon_\ell$ in~\eqref{eq:Energy} are governed by
\begin{equation}
\begin{aligned}
\label{eq:sausage-ms-hamiltonian}
    &\widehat H^{\mathcal{M}}_V \Psi_\ell^{\mathcal{M}}=\varepsilon_\ell\Psi_\ell^{\mathcal{M}} , 
    \qquad \ell \in \mathbb{N},\\
    &\widehat H^{\mathcal{M}}_V
    =
    -\frac{1}{2}\Delta
    +\frac{1}{8}R
    +\frac{\nu\omega_W^2}{4}\frac{x(1-x)}{\mathfrak{t}-x}\,,
\end{aligned}
\end{equation}
where $\Delta$ and $R$ are respectively the Laplace--Beltrami operator and scalar curvature of the metric~\eqref{eq:sausage_metric_rational}. To obtain the dependence of $\varepsilon_\ell$ on the cylinder size $L$, we follow the discussion below~\eqref{eq:Energy} and let the sausage background run along its 1-loop RG flow, evaluated at the scale $L^{-1}$. Concretely, this amounts to the replacement $\mathfrak{t}=(1-(\Lambda_0 L)^{\nu/\pi})^{-1}$, where $\Lambda_0$ is the energy scale dynamically generated by the RG-flow. Since $y$ is a cyclic coordinate, we may restrict to a fixed-momentum sector,
\begin{equation}
    \Psi_\ell^{\mathcal{M}}(x,y)=\re^{\ri \omega_M y}\Psi_\ell(x)\,,
    \qquad \omega_M\in\mathbb{Z}\,.
\end{equation}
Upon making the wavefunction redefinition $\Psi_{\ell}(x) = x^{-\frac{1}{2}}\br{1-x}^{-\frac{1}{2}}\phi_{\ell}(x)$, we further reduce Eq.~\eqref{eq:sausage-ms-hamiltonian} to the normal-form Heun equation
\begin{equation}
\label{eq: Heun t}
\begin{aligned}
\phi_{\ell}''(x)
+\bigg(&
\frac{1-\omega^2_M}{4x^2(x-1)^2}
+\frac{1-\omega_W^2}{4(x-\mathfrak{t})^2}\\
&+\frac{\kappa_{\ell}-12\mathfrak{t}+6}{24x(x-1)(x-\mathfrak{t})}
\bigg)\phi_{\ell}(x)=0\,,
\end{aligned}
\end{equation}
where the accessory parameter $\kappa_\ell$ is related to $\varepsilon_\ell$ by
\begin{equation}
\label{eq:epsilon_kappa}
    \varepsilon_\ell=\frac{\nu}{24}\kappa_\ell\,.
\end{equation}
Eq.~\eqref{eq: Heun t} has four regular singular points at $\{0, 1, \mathfrak{t}, \infty\}$. The physical interval is $x\in(0,1)$, while the movable singularity lies at $\mathfrak{t}>1$; the limits $\mathfrak{t}\to1^{+}$ and $\mathfrak{t}\to\infty$ both correspond to hypergeometric degenerations.

The spectrum is obtained by imposing regularity at both tips of the sausage, which selects the local behaviors
\begin{equation}
\label{eq:heun-boundary-conditions}
\phi_{\ell}(x)
\underset{x\,\to\, 0^+}{\sim}
x^{\frac{1}{2}\left(1+|\omega_M|\right)},
\quad
\phi_{\ell}(x)
\underset{x\,\to\, 1^-}{\sim}
(1-x)^{\frac{1}{2}\left(1+|\omega_M|\right)}.
\end{equation}
To solve the spectral problem we exploit the CFT realization of~\eqref{eq: Heun t} as the classical BPZ equation for a five-point Virasoro block with one degenerate insertion. The AGT representation of Virasoro blocks in terms of Nekrasov partition functions~\cite{Alday:2009aq}, together with the Heun connection formulae of~\cite{Bonelli:2022ten}, allows us to derive exact quantisation conditions for the accessory parameter $\kappa_\ell$ by imposing the local behavior~\eqref{eq:heun-boundary-conditions} --- see Appendix~\ref{app:agt} for details. In this language, the accessory parameter is given by
\begin{equation}
\label{eq:kappa-f-sphere}
\frac{\kappa_\ell}{24} = -\frac{1}{4}+\frac{\mathfrak{t}}{2} + \mathfrak{t}(\mathfrak{t}-1)\partial_{\mathfrak{t}} f(\alpha_\ell,\{\alpha_i\};\mathfrak{t}),
\end{equation}
where $f(\alpha_\ell,\{\alpha_i\};\mathfrak{t})$ is the classical four-point Virasoro conformal block on the sphere defined in~\eqref{eq:cl-four-point-t}. Its external classical momenta $\{\alpha_i\}$ are determined by the local monodromy data of the Heun equation,
\begin{equation}
\label{eq:cft_dictionary}
    \alpha_0=\alpha_1=\frac{1}{2}(1+|\omega_M|)\,,
    \qquad
    \alpha_{\mathfrak t}=\alpha_\infty=\frac{1}{2}(1+|\omega_W|)\,,
\end{equation}
while the internal classical momentum $\alpha_\ell$ is fixed by the boundary conditions~\eqref{eq:heun-boundary-conditions}. Imposing the prescribed behavior at one endpoint of the physical interval selects a local solution, while requiring the same solution to have the prescribed behavior at the other endpoint forces the relevant connection coefficient to vanish:
\begin{equation}
\label{eq:heun-qc-t-frame}
    \Big[
    \Gamma(1-\alpha_\ell+|\omega_M|)
    \Gamma(\alpha_\ell+|\omega_M|)
    \Big]^{-1}=0.
\end{equation}
This condition quantizes the internal momentum $\alpha_\ell$, see Appendix~\ref{app:agt} and Eq.~\eqref{eq:effe}. Together, Eqs.~\eqref{eq:epsilon_kappa},~\eqref{eq:kappa-f-sphere} and~\eqref{eq:heun-qc-t-frame} determine the spectrum $\varepsilon_\ell$ --- see Eq. \eqref{eq:Larget}.

The representation~\eqref{eq:kappa-f-sphere} is naturally adapted to the expansion of the classical block $f(\alpha,\{\alpha_i\};\mathfrak t)$ around $\mathfrak t=\infty$, and hence to the small-$\eta$ regime of the sausage model. It becomes inefficient near the UV limit, where $\eta\to\infty$ and $\mathfrak t\to1^+$. To access this complementary regime it is convenient to pass to the S-dual frame obtained by the change of variables $w=1-x$ and by using $\mathfrak q = 1-\mathfrak t$ as the expansion parameter. The Heun equation~\eqref{eq: Heun t} and the relation~\eqref{eq:kappa-f-sphere} are then rewritten in terms of the new variables and the dual block $g(\alpha,\{\alpha_i\};\mathfrak q)$. The local monodromy data are identified with the classical external momenta as in~\eqref{eq:cft_dictionary}, only with $\alpha_{\mathfrak{t}}$ replaced by $\alpha_{\mathfrak{q}}$, and the accessory parameter is given by
\begin{equation}
\label{eq:kappa-f-sphere-q}
\frac{\kappa_\ell}{24} = \frac{1}{4}-\frac{\mathfrak{q}}{2} - \mathfrak{q}(\mathfrak{q}-1)\partial_{\mathfrak{q}} g(\alpha_\ell,\{\alpha_i\};\mathfrak q).
\end{equation}
The two descriptions use the same universal classical sphere block in two different duality frames, differing in the identification of the cross-ratio and external momenta with the parameters of the Heun equation, see Appendix~\ref{app:agt}. Thus, the two resulting expansions provide complementary analytic descriptions of the same spectrum in the two different regimes.

In the S-dual frame, for $\alpha_{\ell} \neq \frac{1}{2}$ and $|\omega_W|<|\omega_M|+1$, the quantisation condition~\eqref{eq:quant_cond_general} takes the form \footnote{Eq.~\eqref{eq:heun_qc_ns_form} is precisely in the form of a NS quantisation condition~\cite{Nekrasov:2009rc,Nekrasov_2011}, with $\mathcal{W}$ identified as the effective twisted superpotential of the $\mathcal{N}=2$ $\mathrm{SU}(2)$ $N_f = 4$ gauge theory. In the present case, this condition is derived directly from the Heun connection formulae.}
\begin{equation}
\label{eq:heun_qc_ns_form}
\ri \partial_{\alpha_\ell} \mathcal{W}(\alpha_\ell,\{\alpha_i\};\mathfrak q)
=
2\pi\left(\ell+\frac{1}{2}\right),
\qquad \ell \in \mathbb{Z}_{\geq 0},
\end{equation}
where $\mathcal{W}$ is obtained by adding to the classical block $g$ the NS limit of the one-loop contribution to the Nekrasov partition function of the $N_f=4$ $\mathrm{SU}(2)$ gauge theory as in~\eqref{eq:W_def}.
For $|\omega_W|>|\omega_M|+1$, the quantization condition~\eqref{eq:quant_cond_general} also describes eigenfunctions deforming bound states of the UV cigar geometry \cite{Dijkgraaf:1991ba}, see Appendix~\ref{app:agt}.

\begin{figure*}[t]
    \centering
    \includegraphics[width=\textwidth]{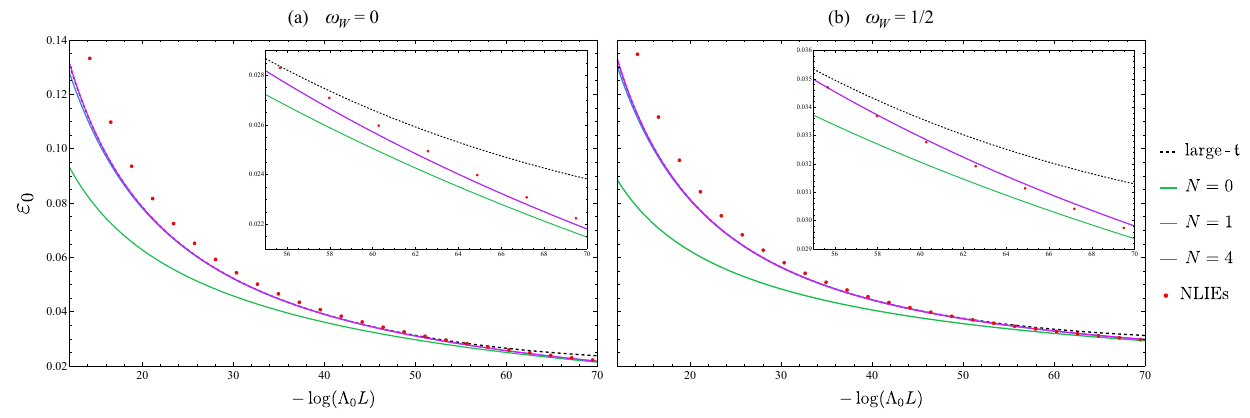}
    \caption{Ground-state point-particle eigenvalue $\varepsilon_0$ for $\nu=4\pi/60$, $\omega_M=0$ and two different values of $\omega_W$. The black dashed curve is obtained from the quantisation condition in~\eqref{eq:heun-qc-t-frame}, using the large-$\mathfrak{t}$ expansion up to order $\mathfrak{t}^{-50}$. The solid curves are obtained from the quantisation condition in~\eqref{eq:heun_qc_ns_form}, with the classical block expanded up to order $\mathfrak{q}^{N}$ for $N=0,1,4$. In this case, the quantisation condition is solved numerically for $\alpha$ after the $\mathfrak{q}$-expansion, allowing us to reach beyond the convergence domain of the double expansion. The red points show the numerical results of the NLIEs of~\cite{Bazhanov:2017nzh}. The inset shows a magnified view of the large-$|\log(\Lambda_0 L)|$ region, where the purple curve completely overlaps the blue one.}
    \label{fig:combined-winding}
\end{figure*}

Solving~\eqref{eq:heun_qc_ns_form} perturbatively, we obtain analytic expressions for the solutions as double expansions,
\begin{equation}
\label{eq:double-expansion-alpha}
    \alpha_\ell = \frac{1}{2}+\sum_{m \geq 0}\alpha_\ell^{(m)}(T)\mathfrak{q}^{m},
\end{equation}
where each coefficient $\alpha_\ell^{(m)}(T)$ is itself an expansion in inverse powers of $T = \frac{1}{4}\left(r_0 - \log \mathfrak{q}\right)$, with $r_0$ defined in~\eqref{eq:r0}. This in turn leads to corresponding double expansions for the accessory parameter $\kappa_\ell$ and hence for the eigenvalue $\varepsilon_\ell$ --- see Eq. \eqref{eq:FOZ_check_2}. In terms of the parameter $\eta=\frac{1}{4}\log(1-\mathfrak{q}^{-1})$ in~\eqref{eq:sausage_metric_trig}, these become expansions in powers of $\eta^{-1}$ and $\re^{-4\eta}$. In this way, the perturbative solution of the quantisation condition gives analytic control on two distinct geometric features of the model. The first is the cylindrical limit of the sausage, in which the $\eta^{-1}$-expansion reflects the approximately quantized momentum along its long direction of length $\sim 2^{3/2}\nu^{-1/2}\eta$. The second is the exponentially small corrections that encode deviations from this limiting geometry, consistent with the fact that the metric depends on $e^{-4\eta}$, see Eq.~\eqref{eq:sausage_metric_trig}.

The same analysis can be carried out in the zero-$\omega_M$, zero-$\omega_W$ sector using the Lamé connection formulae recently derived in~\cite{Bonelli:2025bmt}. Furthermore, the same techniques can be applied to a model~\cite{Fateev:2003sq} obtained by analytic continuation of the sausage metric, see~\cite{Hitchin} for details on both applications. As a first consistency check, we can compare Eq.~\eqref{eq:heun_qc_ns_form} with the semiclassical regime $p_1/n\ll 1$ of the UV quantisation condition in~\cite[Eq.~(5.10)]{Bazhanov:2017nzh}, obtained by gluing together two cigar CFTs. This is recovered from~\eqref{eq:heun_qc_ns_form} by retaining only the $\log\mathfrak{q}$ term in the $\mathfrak{q}$-expansion of $g(\alpha,\{\alpha_i\};\mathfrak{q})$, and identifying
\begin{equation}
\begin{aligned}
&\alpha = \frac{1}{2}+\ri p_1, \quad \omega_W = 2p_2, \quad \nu = \frac{4\pi}{n}, \quad L = R, \\
&\log |\mathfrak{q}| \simeq \frac{4}{n}\log\br{\frac{M R}{2 n}}, \qquad
\Lambda_0 = \frac{\nu M}{8\pi}.
\end{aligned}
\end{equation}
Thus, Eq.~\eqref{eq:heun_qc_ns_form} completes~\cite[Eq.~(5.10)]{Bazhanov:2017nzh} in the semiclassical regime by taking into account the exponentially small corrections to the cylindrical limit of the sausage metric. From the gauge theory viewpoint, these correspond to instanton corrections to the NS twisted effective superpotential $\mathcal{W}$, see Appendix~\ref{app:agt}. Finally, we use the integrability of the sausage model to test our analytic results. In~\cite[Eq.~(5.30)]{Bazhanov:2017nzh}, nonlinear integral equations (NLIEs) were formulated for the exact finite-volume ground state energy $E_0(L)$, at $\omega_M=0$. We solved them numerically \footnote{We warmly thank V. Bazhanov, G. Kotousov and S. Lukyanov for sharing their code solving the NLIEs of~\cite{Bazhanov:2017nzh}.} and, in the regime of validity of the one-loop point-particle approximation, namely $\nu\ll 1$ and $\eta$ sufficiently far from zero, we found excellent agreement with our predictions, see Fig.~\ref{fig:combined-winding}. In particular, the $\re^{-4\eta}$-corrections significantly improve this agreement.

\begin{acknowledgments}

We would like to thank K. Costello, A. Grassi, L. Griguolo, S. Komatsu, G. Kotousov, J. Teschner, E. Tonni, A. Tseytlin and A. Wallberg for useful discussions and comments. We are grateful to V. Bazhanov, G. Kotousov and  S. Lukyanov for sharing their program solving the sausage NLIEs.

The research of T.P. is partially supported by the Swiss National Science Foundation Grant No. 218510. T.P. gratefully acknowledges the hospitality of CERN during the completion of this work.
A.T. is grateful to the F\'ed\'eration de Recherche Interactions Fondamentales (FRIF) for support during a visit at LPTHE in Fall 2025.
The research of G.B. is partly supported by the INFN Iniziativa Specifica ST\&FI and by the PRIN project “Non-perturbative Aspects Of Gauge Theories And Strings”. The research of  T.P. and A.T. is partly supported by the INFN Iniziativa Specifica GAST and Indam GNFM. The research is partly supported by the MIUR PRIN Grant 2020KR4KN2 “String Theory as a bridge between Gauge Theories and Quantum Gravity”. G.B., T.P. and A.T. acknowledge funding from the EU project Caligola (HORIZON-MSCA-2021-SE-01), Project ID: 101086123, and CA21109 - COST Action CaLISTA.
\end{acknowledgments}

\bibliography{BibLetter}

\appendix

\section{iNLSMs}
\label{app:NLSM}

In this Appendix, we describe the basic properties of the iNLSMs that we study in this Letter, including the simplest example of the sausage. We will discuss these models in more detail in future work~\cite{Hitchin}.\\

\paragraph{Twist 1-form and origin.} These iNLSMs can be built from 4d CS theory starting from the \enquote{twist 1-form} $\omega=\omega(z)\dd z$, which is a meromorphic 1-form on $\mathbb{CP}^1$. We take $\omega$ to have $n+3$ real simple poles $(z_r)_{r=1,\dots,n+3}$ and $n+1$ real simple zeroes $(\zeta_i)_{i=1,\dots,n+1}$. Up to M\"{o}bius transformations, these poles and zeroes and an overall factor characterise $\omega$ through $2n+2$ \enquote{twist parameters}, which are the main coupling constants of the associated iNLSM. We further associate a sign $s_i=\pm$ to each zero $\zeta_i$, and accordingly refer to it as chiral or anti-chiral. The 4d CS theory is defined by the action
\begin{equation}\label{eq:CS}
    S_{4d}[A] = \int_{\Sigma \times \mathbb{CP}^{1}} \omega \wedge \text{CS}[A]\,,
\end{equation}
where $\Sigma$ is the NLSM worldsheet with coordinates $(\tau,\s)\sim(\tau,\sigma+L)$, $A$ is an $\slc$-valued gauge field and $\text{CS}[A] = \Tr\bigl(A\wedge\dd A + \frac{2}{3} A\wedge A\wedge A\bigr)$ is the CS 3-form. The gauge field is allowed to have singularities at the zeroes of $\omega$. More precisely, the light-cone component $A_\pm$ has simple poles at the chiral/anti-chiral zeroes $\zeta_i$, for which $s_i=\pm$. Furthermore, $A$ must satisfy some boundary conditions at the poles $(z_r)_{r=1,\dots,n+3}$ of $\omega$: its evaluation at these points belongs to an isotropic subalgebra of the \enquote{defect Lie algebra} $\slc^{\times(n+3)}$. The specific iNLSMs that we consider in this Letter correspond to the case where this subalgebra is formed by triangular matrices, with appropriate relations among the $n+3$ diagonal elements. These relations depend on the choice of additional \enquote{isotropy parameters}, which will also enter in the NLSM, although they play a less important role than the twist parameters defining $\omega$.\\

\paragraph{Reduction to 2d.} To reduce the 4d theory to an integrable 2d NLSM, we first perform a gauge transformation $A \mapsto \Lc=\Lc_\tau\dd\tau + \Lc_\sigma\dd \sigma$ killing the components of the gauge field along $\mathbb{CP}^1$. The equations of motion of the theory then become
\begin{equation}
    \p_{\overline{z}}\Lc = 0\,, \qquad F_\Sigma(\Lc)=0\,,
\end{equation}
away from the zeroes of $\omega$, with $F_\Sigma(\Lc)$ the curvature of $\Lc$ as a 2d connection on $\Sigma$. The first equation implies that $\Lc_\pm=\Lc_\pm(z)$ is a meromorphic function of $z$, with simple poles at the chiral/anti-chiral zeroes of $\omega$, while the second implies that $\Lc(z)$ is flat along $\Sigma$. These are exactly the defining properties of the Lax connection of an integrable 2d field theory on the worldsheet $\Sigma$, with spectral parameter $z$. The fundamental fields $(X^\mu(\tau,\sigma))_{\mu=1,\dots,d}$ of this theory arise as edge modes when performing the gauge transformation $A\mapsto\Lc$, due to the boundary conditions imposed on the gauge field at the poles of $\omega$. Imposing these boundary conditions further expresses the Lax connection $\Lc_\pm(z)$ in terms of the fields $X^\mu$ and their derivatives $\p_\pm X^\mu$. Finally, integrating over $z\in\mathbb{CP}^1$ in the 4d CS action in~\eqref{eq:CS} gives the 2d action $S[X]$ given in~\eqref{eq:NLSM}. This allows us to extract the metric $G_{\mu\nu}(X)$ and B-field $B_{\mu\nu}(X)$ of the NLSM target space $\Mc$, in terms of the twist and isotropy parameters.

Our choice of isotropic subalgebra leads to the existence of $d-n$ commuting isometries, with adapted coordinates $(X^a)_{a=n+1,\dots,d}$. In these coordinates, the metric and B-field depend only on $(X^r)_{r=1,\dots,n}$. The geometry $\Mc$ is thus locally fibred over a base $\Bc$, with base and fibre coordinates $X^r$ and $X^a$, respectively. The base geometry depends only on the twist parameters, while the fibre geometry also involves the isotropy parameters. This description is local: the global geometry and topology, including fibre compactness and possible collapse to a point, depend on the details of the construction.

This iNLSM also admits an affine Gaudin Hamiltonian description~\cite{Feigin:2007mr,Vicedo:2017cge,Delduc:2019bcl}, where $\omega$ controls the Poisson bracket of $\Lc_\sigma(z)$, ensuring Poisson-commutativity of the conserved monodromy charges. Moreover, the model is 1-loop renormalisable, with an RG-flow that can be written compactly in terms of $\omega$~\cite{Delduc:2020vxy,Derryberry:2021rne,Lacroix:2025ias,Costello:2026gve}. Geometrically, this means that the corresponding background $(\Mc,G,B)$ is stable under the torsionful Ricci flow.

As explained in the main text, in this Letter we conjecturally identify the NLSM's 1-loop particle Hamiltonian with the Perelman operator associated to this Ricci flow and a quantum Hitchin Hamiltonian. Unlike the twist parameters, the isotropy parameters do not run under the RG-flow and enter the Hitchin correspondence only through the dictionary relating parabolic weights to momentum/winding parameters along the fibres.\\

\paragraph{Reducing 4d CS theory.}
We now sketch the implementation of the classical point-particle ansatz in 4d CS theory, starting by restricting to fields independent of the spatial coordinate $\sigma\sim\sigma+L$. Decomposing the gauge field as $A=a+\phi\,\dd\sigma$, with
\begin{equation}
    a=A_\tau\,\dd\tau+A_{\bar z}\,\dd\bar z,
    \qquad
    F(a)=\dd a+a\wedge a,
\end{equation}
and integrating over the circle in the 4d CS action in~\eqref{eq:CS} gives
\begin{equation}\label{3dcs}
 S_{\mathrm{red}}[a,\phi] = L\int_{\mathbb R\times \mathbb{CP}^1}\omega\wedge
 \left[ 2\Tr\bigl(\phi F(a)\bigr) -\dd\Tr\bigl(\phi a\bigr) \right]\,.
\end{equation}
Away from the poles of $\omega$, the second term is a total derivative, while the bulk term defines a topological-holomorphic BF theory with auxiliary field $\Phi=L\,\phi\omega$. Performing a gauge transformation $a\mapsto \Lc_\tau\,\dd\tau$ to eliminate the $\dd\bar z$-component of the gauge field, the bulk equations of motion become
\begin{equation}
    \partial_{\bar{z}} \Lc_\tau = \partial_{\bar{z}} \Phi = 0, \qquad \partial_{\tau}\Phi = \left[\Phi, \Lc_\tau \right]\,.
\end{equation}
The first equation implies that $\Lc_\tau$ and $\Phi$ are meromorphic in $z$, with singularities at the zeroes and poles of $\omega$ respectively. The last equation has Lax form, with $\Phi(z)$ as the Lax matrix and $\Lc_\tau(z)$ as the auxiliary matrix, and therefore describes an isospectral evolution. More precisely, using the results of~\cite{Vicedo:2022mrm}, we expect this integrable mechanical model to be a Hitchin system, with $\Phi(z)$ identified with the Higgs field, in agreement with the results of Section~\ref{sec:iNLSM}. The full identification would require a careful treatment of defect terms in the BF-theory~\eqref{3dcs} at the poles and zeroes of $\omega$, which we expect to be related to the A- and B-defects of~\cite{Vicedo:2022mrm}. We leave these investigations for future work.\\

\paragraph{Sausage example.} 
A particularly simple example is the sausage model~\cite{Fateev:1992tk}, obtained from the above construction by imposing an additional $\mathbb{Z}_2$-equivariance condition~\cite{Costello:2019tri,Schmidtt:2020dbf} on the gauge field under $z\mapsto -z$ on $\mathbb{CP}^1$ and an involutive automorphism of $\slc$. We  focus here on the resulting 2d NLSM, whose target-space metric is~\cite{Fateev:1992tk}
\begin{equation}
\label{eq:sausage_metric_trig}
    \dd s^2 = \frac{2\sinh(2\eta)}{\nu} \frac{\dd Z^2 + \dd Y^2}{\cosh(2\eta)+\cosh(2Z)}\,,
\end{equation}
where $\nu,\eta$ are positive parameters, and $Z\in(-\infty,+\infty)$, $Y\sim Y+2\pi$ are the coordinates. We can introduce
\begin{equation}
\label{eq:rational_variables}
    X = \left( 1 + \re^{-2(Z+\eta)} \right)^{-1}, \qquad \mathfrak{t} = \left( 1 - \re^{-4\eta} \right)^{-1}\,,
\end{equation}
with $X\in(0,1)$ and $\mathfrak{t}>1$, obtaining:
\begin{equation}
    \dd s^2 = \frac{1}{2\nu(\mathfrak{t}-X)} \left( \frac{\dd X^2}{X(1-X)} + 4 X(1-X)\dd Y^2 \right)\,. 
\end{equation}
In the language of the previous subsections, $X$ is the base coordinate while $Y$ describes a compact isometric fibre. The metric degenerates as $X \to 0,1$, where a change of chart is required to describe the local geometry.   The twist 1-form of the sausage model is given by~\cite{Delduc:2013fga}
\begin{equation}
    \omega = \frac{\re^{4\eta}-1}{\nu} \frac{2z}{(z^2-1)(z^2-e^{4\eta})}\dd z\,.
\end{equation}
It has 4 simple poles at $z_{1,2}=\pm 1$ and $z_{3,4}=\pm \re^{2\eta}$, as well as two simple zeroes at $z=\infty$ and $z=0$, which we take to be respectively chiral and anti-chiral. Reflecting the $\mathbb{Z}_2$-equivariance under $z\mapsto-z$, the poles occur in opposite pairs, while the zeroes lie at the fixed points. In this example there are no free isotropy parameters in the choice of boundary conditions of the gauge field, so that the twist parameters $(\nu,\eta)$ -- or $(\nu,\mathfrak{t})$ -- are the only coupling constants of the NLSM. The classical equations of motion can be recast as the flatness of a $\mathfrak{sl}(2,\mathbb{C})$-valued Lax connection whose matrix entries are
\begin{subequations}\label{eq:LaxSausage}
\begin{align}
[\Lc_+(z)]_{11}
 &= -[\Lc_+(z)]_{22} = \frac{(X^2-2X+\mathfrak{t})\p_+X}
          {4X(X-1)(X-\mathfrak{t})} \notag\\
 &\quad-\frac{\ri(X^2-2\mathfrak{t}X+\mathfrak{t})\p_+Y}
                  {2(X-\mathfrak{t})},\\
[\Lc_+(z)]_{12}
 &= \frac{z(\p_+X-2\ri X(X-1)\p_+Y)}{2X},\\
[\Lc_+(z)]_{21}
 &= \frac{z(\mathfrak{t}-1)(\p_+X+2\ri X(X-1)\p_+Y)}
          {2(X-1)(\mathfrak{t}-X)},\\
[\Lc_-(z)]_{11}
 &= -[\Lc_-(z)]_{22}=-\frac{(X^2-\mathfrak{t})\p_-X}
           {4X(X-1)(X-\mathfrak{t})} \notag\\
 &\quad-\frac{\ri(X^2-2\mathfrak{t}X+\mathfrak{t})\p_-Y}
                  {2(X-\mathfrak{t})},\\
[\Lc_-(z)]_{12}
 &= \frac{\mathfrak{t}(\p_-X-2\ri X(X-1)\p_-Y)}
          {2zX(\mathfrak{t}-X)},\\
[\Lc_-(z)]_{21}
 &= \frac{\p_-X+2\ri X(X-1)\p_-Y}{2z(X-1)}.
\end{align}
\end{subequations}
As expected, $\Lc_\pm(z)$ have poles at $\infty$ and $0$, respectively. The boundary conditions make $\Lc_\pm(z_r)$ triangular after a field-dependent gauge transformation, while equivariance relates $\Lc_\pm(z)$ and $\Lc_\pm(-z)$ by a $\slc$-automorphism.

Using the point-particle ansatz in~\eqref{eq:minisuperspace_ansatz}, with $\omega_W$ encoding fibre winding/quasi-periodicity and the conserved momentum conjugate to $y$ fixed to $\omega_M$, we obtain
\begin{align}
& H = \frac{p^2}{2G_{xx}(x)}
   + \frac{\omega_M^2}{2G_{yy}(x)}
   + \frac{\nu^2}{8}G_{yy}(x)\omega_W^2 \label{eq:SausageH}\\
  &= \frac{\nu}{4}\left[
       4x(1-x)(\mathfrak{t}-x)p^2
       + \frac{\mathfrak{t}-x}{x(1-x)}\omega_M^2
       + \frac{x(1-x)}{\mathfrak{t}-x}\omega_W^2
     \right],\notag
\end{align}
where $p$ denotes the momentum conjugate to $x$. This is the Hamiltonian of a $\mathbb{Z}_2$-equivariant Hitchin system with 4 punctures at the poles $z_r$ of $\omega$. We identify the 1-form $\Phi=L\,\omega(z)\Lc_\sigma(z)\dd z$ in the point-particle ansatz with the Hitchin Higgs field. Using the NLSM Lax connection in~\eqref{eq:LaxSausage}, we find
\begin{widetext}
\begin{align}
    \Phi =& \frac{1}{2(z^2-1)((\mathfrak{t}-1)z^2-\mathfrak{t})} \\
    &\hspace{10pt}\begin{pmatrix}
        z\frac{2x(x-1)(\mathfrak{t}-x)p+\ri(x^2-2\mathfrak{t}x+\mathfrak{t})\omega_W}{\mathfrak{t}-x} & \frac{((x-\mathfrak{t})z^2+\mathfrak{t})(2x(x-1)p-\ri \omega_M)}{x} - \frac{\ri (x-1)((\mathfrak{t}-x)z^2+\mathfrak{t})\omega_W}{\mathfrak{t}-x}  \\[3pt]
        \frac{((\mathfrak{t}-1)z^2-\mathfrak{t}+x)(2x(x-1)p+\ri\omega_M)}{1-x} + \frac{\ri x((\mathfrak{t}-1)z^2+\mathfrak{t}-x)\omega_W}{\mathfrak{t}-x} & -z\frac{2x(x-1)(\mathfrak{t}-x)p+\ri(x^2-2\mathfrak{t}x+\mathfrak{t})\omega_W}{\mathfrak{t}-x}
    \end{pmatrix}\,\dd z\,. \notag
\end{align}
\end{widetext}
\par\null\clearpage\noindent
The quadratic differential of the Hitchin system is obtained as $Q(z)\dd z^2=\frac{1}{2}\Tr(\Phi^2)$, which gives
{\small\begin{align}
    &4Q(z) = -\frac{\omega_-^2}{(z-z_1)^2}-\frac{\omega_-^2}{(z-z_2)^2} - \frac{\omega_+^2}{(z-z_3)^2} - \frac{\omega_+^2}{(z-z_4)^2} \notag \\ 
    &+ \left((1-2\mathfrak{t}) \omega_+^2 +2(1-\mathfrak{t})\omega_-^2 + \frac{2H}{\nu} \right) \left( \frac{z_1^{-1}}{z-z_1} + \frac{z_2^{-1}}{z-z_2} \right)\notag\\ 
    &+ \left(2\mathfrak{t}\, \omega_+^2 +(2\mathfrak{t}-1)\omega_-^2 - \frac{2H}{\nu} \right) \left( \frac{z_3^{-1}}{z-z_3} + \frac{z_4^{-1}}{z-z_4} \right)\,,
\end{align}}where $H$ is the particle Hamiltonian in~\eqref{eq:SausageH} and $\omega_\pm=(\omega_M\pm\omega_W)/2$. 
The constant double-pole coefficients $-\omega_\pm^2/4$ at $z=z_r$ are characteristic of a Hitchin quadratic differential and relate its parabolic weights to the momentum and winding numbers $\omega_{M/W}$, as expected. Moreover, $Q(z)$ encodes the particle Hamiltonian $H$, which can be extracted as
\begin{equation}
    H = \mathop{\mathrm{res}}_{z=\infty} \frac{Q(z)}{\omega(z)}\dd z - \mathop{\mathrm{res}}_{z=0} \frac{Q(z)}{\omega(z)}\dd z\,,
\end{equation}
in agreement with the general formula in~\eqref{eq:HHitchin}.

For this low-dimensional Hitchin system, it follows immediately that $\lbrace Q(z),Q(w)\rbrace=0$ since $Q(z)$ depends only on $H$. Nevertheless, $\Phi(z)$ retains non-trivial integrable structures: it satisfies an $\mathcal{R}$-matrix Poisson bracket and evolves isospectrally under the Hamiltonian flow of $Q(w)$, \textit{i.e.} through a commutator.

\section{Conformal blocks, Nekrasov functions \& AGT dictionary}
\label{app:agt}
In this Appendix, we briefly review the 2d CFT approach to the Heun connection problem developed in \cite{Bonelli:2022ten}. We write the central charge as $c = 1+6Q^2$, $Q = b + b^{-1}$ and consider the four-point Virasoro conformal blocks
\begin{subequations}
\label{eq:4point}
\begin{align}
\mathcal{F}\br{A,\{A_j\};\mathfrak{t}}
& =
\left\langle
V_{A_\infty}(\infty)V_{A_{\mathfrak{t}}}(\mathfrak{t})V_{A_1}(1)V_{A_0}(0)
\right\rangle, \label{eq:4pointBt} \\
\mathcal{G}\br{A,\{A_j\};\mathfrak{q}}
&=
\left\langle
V_{A_\infty}(\infty)V_{A_1}(1)V_{A_{\mathfrak{q}}}(\mathfrak{q})V_{A_0}(0)
\right\rangle, \label{eq:4pointBq}
\end{align}
\end{subequations}
where $A_j$ parametrizes the holomorphic conformal dimension of the corresponding chiral vertex operator according to $\Delta_{A_j} = A_j(Q-A_j)$. The block in~\eqref{eq:4pointBt} is understood to be in the OPE channel in which $V_{A_{\mathfrak{t}}}$ is fused with $V_{A_\infty}$, and thus admits a large-$\mathfrak{t}$ expansion. Meanwhile, the block in~\eqref{eq:4pointBq} is understood to be in the OPE channel in which $V_{A_{\mathfrak{q}}}$ is fused with $V_{A_0}$, and thus admits a small-$\mathfrak{q}$ expansion. In each case, $A$ labels the internal dimension in the chosen channel, with $\Delta_A = A(Q-A)$. In the classical limit, $b\to0$ with $A_j=\alpha_j/b$, $A=\alpha/b$ and $\alpha_j$, $\alpha$ held fixed, we define the classical four-point blocks
\begin{subequations}
\begin{align}
f\br{\alpha,\{\alpha_j\};\mathfrak{t}}
& =
\lim_{b\to0}
b^2\log\mathcal{F}\br{A,\{A_j\};\mathfrak{t}},\label{eq:cl-four-point-t}\\
g\br{\alpha,\{\alpha_j\};\mathfrak{q}}
&=
\lim_{b\to0}
b^2\log\mathcal{G}\br{A,\{A_j\};\mathfrak{q}}.\label{eq:cl-four-point-q}
\end{align}
\end{subequations}
We introduce the classical fusion and braiding matrices
\begin{subequations}
\begin{align}\label{eq:effe}
& \mathbb{F}_{rs} \big(\tilde{\theta}_1,\tilde{\theta}_2,\tilde{\theta}_3\big)
=
\frac{
\Gamma\br{1+r\tilde{\theta}_1}
\Gamma\br{-s\tilde{\theta}_2}
}{
\prod_{\sigma=\pm1}
\Gamma\br{
\frac{1+r\tilde{\theta}_1-s\tilde{\theta}_2+\sigma\tilde{\theta}_3}{2}}
},\\
& \mathbb{B}_{rs} \big(\tilde{\theta}_1,\tilde{\theta}_2,\tilde{\theta}_3\big)
=
\frac{\re^{-\frac{\mathrm{i}\pi}{2}
\bigl(1+r\tilde{\theta}_1+\tilde{\theta}_2-s\tilde{\theta}_3\bigr)}
\Gamma\br{1+r\tilde{\theta}_1}
\Gamma\br{s\tilde{\theta}_3}
}{
\prod_{\sigma=\pm 1}\Gamma\br{\frac{1+s\tilde{\theta}_3+r\br{\tilde{\theta}_1+\sigma\tilde{\theta}_2}}{2}}
},
\end{align}
\end{subequations}
where $r,s =\pm 1$. For generic values of the parameters $\{\alpha_j\}$ in~\eqref{eq:cft_dictionary}, we consider a basis of Frobenius solutions at $x=0$ and $x=1$ of Eq.~\eqref{eq: Heun t}, normalized by
\begin{align}
\phi^{(0)}_{r}\br{\alpha,\{\alpha_j\};\mathfrak{t},x} &= x^{\frac{1}{2}+r\br{\alpha_{0}-\frac{1}{2}}}(1+\mathcal{O}(x)),\\
\phi^{(1)}_{r}\br{\alpha,\{\alpha_j\};\mathfrak{t},x} &= (1-x)^{\frac{1}{2}+r\br{\alpha_{1}-\frac{1}{2}}}(1+\mathcal{O}(1-x)),\notag
\end{align}
with $r = \pm 1$. Then, for $|\mathfrak t|>1$, the analytic continuation of the Frobenius basis from $x=0$ to $x=1$ is described by
\begin{align}
&\phi_{r}^{(0)}\br{\alpha,\{\alpha_j\};\mathfrak{t},x} = \sum_{s=\pm}\mathbb{F}_{rs}\br{\theta_0, \theta_1, \theta}\label{eq:connection-t}\\
&\times\re^{\frac{r}{2}\partial_{\alpha_{0}} f(\alpha,\{\alpha_j\};\mathfrak t)-\frac{s}{2}\partial_{\alpha_{1}} f(\alpha,\{\alpha_j\};\mathfrak t)}\phi_{s}^{(1)}\br{\alpha,\{\alpha_j\};\mathfrak{t},x},\notag
\end{align}
where $\theta_j = 2\alpha_j-1$, $\theta = 2\alpha-1$ and the derivatives of the classical block are taken before the specialization to the dictionary in~\eqref{eq:cft_dictionary}. Using this dictionary, the boundary conditions in~\eqref{eq:heun-boundary-conditions} select $\phi_+^{(0)}$ at $x=0$. Requiring the coefficient of $\phi_{-}^{(1)}$ to vanish then implies the quantisation condition in~\eqref{eq:heun-qc-t-frame}, where an overall factor containing $\Gamma(1+|\omega_M|)\Gamma(|\omega_M|)$ has been removed. The resulting condition extends regularly to the resonant values $\omega_M\in\mathbb Z$. 

Under the change of variables $(\mathfrak{t},x) \mapsto (\mathfrak{q}=1-\mathfrak{t},w=1-x)$, the connection formula adapted to $|\mathfrak q|<1$ takes the form
\begin{align}
&\phi_{\epsilon}^{(0)}
 \br{\alpha,\{\alpha_j\};\mathfrak q,w}
=\sum_{r,s=\pm}
 \re^{\ri\pi\alpha_{\mathfrak q}
 +\frac{\epsilon}{2}\partial_{\alpha_0}g
 -\frac{s}{2}\partial_{\alpha_1}g
 +\frac{r}{2}\partial_\alpha g}\notag
\\
&\times
 \mathbb{B}_{\epsilon r}
 \br{\theta_0,\theta_{\mathfrak q},\theta}
 \mathbb{F}_{rs}
 \br{\theta,\theta_1,\theta_\infty}
 \phi_s^{(1)}
 \br{\alpha,\{\alpha_j\};\mathfrak q,w},\label{eq:connection-q}
\end{align}
where $g = g(\alpha,\{\alpha_j\} ; \mathfrak{q})$, $\log \mathfrak{q} = \log|\mathfrak{q}|+\ri \pi$ and with local solutions normalized by
\begin{align}
\phi^{(0)}_{r}\br{\alpha,\{\alpha_j\};\mathfrak{q},w} &= w^{\frac{1}{2}+r\br{\alpha_{0}-\frac{1}{2}}}(1+\mathcal{O}(w)),\\
\phi^{(1)}_{r}\br{\alpha,\{\alpha_j\};\mathfrak{q},w} &= (1-w)^{\frac{1}{2}+r\br{\alpha_{1}-\frac{1}{2}}}(1+\mathcal{O}(1-w)).\notag
\end{align}
The derivatives of the classical block are again taken before the specialization to $\alpha_0=\alpha_1=\frac{1}{2}(1+|\omega_M|)$ and $\alpha_{\mathfrak q}=\alpha_\infty=\frac{1}{2}(1+|\omega_W|)$. As before, through this dictionary the boundary condition at $w=0$ selects $\phi_+^{(0)}$. Requiring the coefficient of $\phi_-^{(1)}$ to vanish then implies the quantisation condition
\begin{equation}
\label{eq:quant_cond_general}
\begin{aligned}
&\re^{\frac12\partial_\alpha g+\ri\pi(\alpha-\frac12)}
\frac{\Gamma(2\alpha-1)\Gamma(2\alpha)}
{\displaystyle\prod_{\sigma=\pm1}
 \Gamma\!\left(\alpha+\frac{|\omega_M|}{2}
 +\sigma\frac{|\omega_W|}{2}\right)^2}
\\
&+ \big(\alpha \to 1-\alpha\big) = 0.
\end{aligned}
\end{equation}
Also in this case an overall factor has been removed, and the resulting condition extends to resonant values $\omega_M\in\mathbb Z$. It therefore applies to all physical values $\omega_M\in\mathbb Z$ and $\omega_W\in\mathbb R$, with the appropriate limit taken whenever necessary.

For $\alpha\neq\tfrac12$ and $|\omega_W|<|\omega_M|+1$, the previous quantisation condition can be rewritten in the form of Eq.~\eqref{eq:heun_qc_ns_form}, where the effective twisted superpotential $\mathcal W$ is defined, up to an $\alpha$-independent term, by
\begin{equation}
\label{eq:W_def}
    \partial_\alpha \mathcal{W} = \partial_\alpha g(\alpha,\{\alpha_j\} ; \mathfrak{q}) +2\pi\ri(\alpha-1)+2\log\mathcal B(\alpha),
\end{equation}
and
\begin{equation}
\label{eq:Galpha}
\mathcal{B}(\alpha)
=
\frac{\Gamma(2\alpha-1)}{\Gamma(1-2\alpha)}
\prod_{\sigma=\pm 1}
\frac{
\Gamma\left(
1-\alpha+\frac{|\omega_M|}{2}
+\sigma\frac{|\omega_W|}{2}
\right)
}{
\Gamma\left(
\alpha+\frac{|\omega_M|}{2}
+\sigma\frac{|\omega_W|}{2}
\right)
}.
\end{equation}
In this parameter range, the logarithm of $\mathcal B(\alpha)$ is taken on the branch for which $\log\mathcal B(\alpha) \to \ri\pi$ as $\alpha\to\frac12$. The constant $r_0$ entering the definition of the variable $T$ in~\eqref{eq:double-expansion-alpha} is given by
\begin{equation}
\label{eq:r0}
\begin{aligned}
r_0 & = \ri \pi -4\gamma_E -2\psi\br{\frac{1+|\omega_M|+|\omega_W|}{2}}\\
& -2\psi\br{\frac{1+|\omega_M|-|\omega_W|}{2}},
\end{aligned}
\end{equation}
where $\psi$ is the digamma function. For $|\omega_W|>|\omega_M|+1$, Eq.~\eqref{eq:quant_cond_general} also describes eigenfunctions deforming bound states of the $\mathfrak q\to0$ hypergeometric degeneration (cigar limit~\cite{Dijkgraaf:1991ba}). These are already included among the levels described by the large-$\mathfrak t$ quantisation condition, but require small-$\mathfrak q$ expansions different from those in~\eqref{eq:double-expansion-alpha}. Threshold and resonant cases require taking the appropriate limit in the full condition~\eqref{eq:quant_cond_general}. A more detailed discussion will be presented in~\cite{Hitchin}.

The AGT correspondence~\cite{Alday:2009aq} provides a combinatorial representation of Virasoro conformal blocks on the sphere in terms of instanton Nekrasov partition functions of linear $\mathrm{SU}(2)$ quiver gauge theories. In the present setting, it
provides combinatorial expressions for the conformal blocks determining both the Frobenius solutions and the accessory parameter of the Heun equation~\cite{Bonelli:2022ten}. Let us define
\cite{Flume:2002az,Bruzzo:2002xf}
\begin{subequations}
\begin{align}
&z_{\mathrm{vec}}(\vec a,\vec Y)
=
\prod_{i,j=1}^{2}\prod_{s\in Y_i}
\frac{1}{
E_{ij}(s)\bigl[\epsilon_1+\epsilon_2-E_{ij}(s)\bigr]
},
\\
&z_{\mathrm f}(\vec a,\vec Y;m)
=
\prod_{i=1}^{2}\prod_{s\in Y_i}
\bigl(\varphi_i(s)-m+\epsilon_1+\epsilon_2\bigr),
\\
&z_{\mathrm{af}}(\vec a,\vec Y;m)
=
\prod_{i=1}^{2}\prod_{s\in Y_i}
\bigl(\varphi_i(s)+m\bigr).
\end{align}
\end{subequations}
Here $\vec a=(a_1,a_2)$ and $\vec Y=(Y_1,Y_2)$ is a pair of Young diagrams. For a box $s=(r,c)\in Y_i$, we also define
\begin{equation}
\begin{aligned}
E_{ij}(s)
&=
a_i-a_j-\epsilon_1L_{Y_j}(s)
+\epsilon_2\bigl(A_{Y_i}(s)+1\bigr),
\\
\varphi_i(s)
&=
a_i+\epsilon_1(r-1)+\epsilon_2(c-1),
\end{aligned}
\end{equation}
where
\begin{equation}
\begin{aligned}
& A_Y(r,c)=\lambda_r-c, \qquad L_Y(r,c)=\lambda_c^{\mathrm T}-r, \\
& Y = \{\lambda_1, \lambda_2, \dots \}, \qquad \lambda_1 \geq \lambda_2 \geq \dots \geq 0.
\end{aligned}
\end{equation}
For the $\mathrm{SU}(2)$ theory with two antifundamental hypermultiplets of masses $m_1,m_2$ and two fundamental hypermultiplets of masses $m_3,m_4$, the instanton partition function reads
\begin{equation}
\begin{aligned}
\mathcal{Z}_{\mathrm{inst}}
&=
\sum_{\vec{Y}}
q^{|\vec{Y}|}
z_{\mathrm{vec}}(\vec{a},\vec{Y})
z_{\mathrm{af}}(\vec{a},\vec{Y};m_1)
z_{\mathrm{af}}(\vec{a},\vec{Y};m_2)\\
& \quad\quad \times z_{\mathrm{f}}(\vec{a},\vec{Y};m_3)
z_{\mathrm{f}}(\vec{a},\vec{Y};m_4).
\end{aligned}
\end{equation}
Here $|\vec{Y}|=|Y_1|+|Y_2|$ and $\vec{a}=(a,-a)$ parametrizes the vacuum expectation value of the adjoint scalar in the $\mathcal{N}=2$ vector multiplet. The parameters $\epsilon_1$ and $\epsilon_2$ are the equivariant parameters of the spacetime torus action and $q$ is the instanton counting parameter. With the identification
\begin{equation}
\begin{aligned}
& \widehat{m}_1 = A_{\mathfrak{t}}+A_\infty-Q/2, \quad \widehat{m}_2 = A_{\mathfrak{t}}-A_\infty+Q/2,\\
& \widehat{m}_3 =A_1+A_0-Q/2, \quad \widehat{m}_4 = A_1-A_0+Q/2, \\
& a/\sqrt{\epsilon_1 \epsilon_2} = A - Q/2, \quad b^2 = \epsilon_2/\epsilon_1, \quad q = \mathfrak{t}^{-1},
\end{aligned}
\end{equation}
where $\widehat{m}_i = m_i/\sqrt{\epsilon_1 \epsilon_2}$, the four-point block $\mathcal{F}\br{A,\{A_i\};\mathfrak{t}}$ is expressed in terms of $\mathcal{Z}_{\mathrm{inst}}$ as
\begin{equation}
\begin{aligned}
\mathcal{F} \br{A,\{A_i\};\mathfrak{t}} & = \mathfrak{t}^{\Delta_{A_\infty}-\Delta_A-\Delta_{A_{\mathfrak{t}}}}\\
& \times (1-\mathfrak{t}^{-1})^{-2A_{\mathfrak{t}}\left(Q-A_1\right)}
\mathcal{Z}_{\mathrm{inst}}.
\end{aligned}
\end{equation}
Similarly, with the identification
\begin{equation}
\begin{aligned}
& \widehat{m}_1 = A_{1}+A_\infty-Q/2, \quad \widehat{m}_2 = A_{1}-A_\infty+Q/2,\\
& \widehat{m}_3 =A_{\mathfrak{q}}+A_0-Q/2, \quad \widehat{m}_4 = A_{\mathfrak{q}}-A_0+Q/2, \\
& a/\sqrt{\epsilon_1 \epsilon_2} = A - Q/2, \quad b^2 = \epsilon_2/\epsilon_1, \quad q = \mathfrak{q},
\end{aligned}
\end{equation}
the four-point block $\mathcal{G}\br{A,\{A_i\};\mathfrak{q}}$ is expressed in terms of $\mathcal{Z}_{\mathrm{inst}}$ as
\begin{equation}
\begin{aligned}
\mathcal{G} \br{A,\{A_i\};\mathfrak{q}} & = \mathfrak{q}^{\Delta_A-\Delta_{A_{\mathfrak{q}}}-\Delta_{A_0}}\\
& \times (1-\mathfrak{q})^{-2A_{1}\left(Q-A_{\mathfrak{q}}\right)}
\mathcal{Z}_{\mathrm{inst}}.
\end{aligned}
\end{equation}
The Frobenius solutions $\phi_{\epsilon}^{(i)}$ also admit a combinatorial representation in terms of Nekrasov functions. Since we do not use this representation explicitly in this Letter, we refer to~\cite{Alday:2009aq,Bonelli:2022ten,Hitchin} for further details. To conclude, the first few terms of the expansion of the classical four-point blocks $f(\alpha, \{\alpha_j\} ; \mathfrak{t})$ and $g(\alpha, \{\alpha_j\} ; \mathfrak{q})$ read, respectively, 
\begin{equation}
\begin{aligned}
f(\alpha,& \{\alpha_j\} ; \mathfrak{t}) = \br{\delta_\infty-\delta-\delta_{\mathfrak t}}\log\mathfrak t \\
& +\frac{\br{\delta+\delta_{\mathfrak t}-\delta_\infty}
\br{\delta+\delta_1-\delta_0}}{2\delta}\mathfrak t^{-1}
+\mathcal O(\mathfrak t^{-2}),
\end{aligned}
\end{equation}
and
\begin{equation}
\begin{aligned}
g(\alpha,& \{\alpha_j\} ; \mathfrak{q}) =\br{\delta-\delta_{0}-\delta_{\mathfrak{q}}} \log \mathfrak{q} \\
& + \frac{\br{\delta-\delta_{0}+\delta_{\mathfrak{q}}}\br{\delta+\delta_{1}-\delta_{\infty}}}{2\delta}\mathfrak{q} + \mathcal{O}(\mathfrak{q}^2),
\end{aligned}
\end{equation}
where $\delta = \alpha(1-\alpha)$ and $\delta_i = \alpha_i (1-\alpha_i)$.

\section{Explicit expansions}
\label{app:expansions}
In this Appendix, we collect examples of the expansions obtained from the quantisation conditions discussed in the main text.
\paragraph{Large-$\mathfrak{t}$ expansions.}
Beginning from the large-$\mathfrak{t}$ expansion of the quantized accessory parameter $\kappa_\ell$, determined by Eq.~\eqref{eq:kappa-f-sphere}, we find
\begin{widetext}
\begin{equation}\label{eq:Larget}
\begin{aligned}
\kappa_\ell\br{\omega_M, \omega_W; \mathfrak{t}}
&= (12-24\delta_\ell)\mathfrak{t}
+ 12\delta_\ell - 6 + \frac{3}{4\delta_\ell+3}
\left(\frac{1}{\mathfrak{t}}+\frac{1}{2\mathfrak{t}^{2}}\right)
\Bigl[
3\delta_\ell^{2}+\delta_\ell-1
+(\delta_\ell+1)\Sigma
-\Pi
\Bigr] \\
&+ \frac{3}{
16(4\delta_\ell+3)^{3}(4\delta_\ell+15)\mathfrak{t}^{3}}
\Biggl\{
972\delta_\ell^{5}
+5425\delta_\ell^{4}
+7378\delta_\ell^{3}
+2331\delta_\ell^{2}
-1337\delta_\ell-687\\
&+2\bigl(
164\delta_\ell^{4}
+1027\delta_\ell^{3}
+1878\delta_\ell^{2}
+1361\delta_\ell+345
\bigr)\Sigma
-\bigl(
4\delta_\ell^{3}
+15\delta_\ell^{2}
+17\delta_\ell+3
\bigr)
\bigl(\Sigma^2-2\Pi\bigr) \\
&-4\bigl(
84\delta_\ell^{3}
+423\delta_\ell^{2}
+494\delta_\ell+165
\bigr)
\Pi
-2\bigl(12\delta_\ell^{2}+13\delta_\ell+15\bigr)
\Pi \,\Sigma -(20\delta_\ell-33)\Pi^2
\Biggr\} +\mathcal{O}\!\left(\mathfrak{t}^{-4}\right),
\end{aligned}
\end{equation}
\end{widetext}
where we use the notation
\begin{equation}
\begin{aligned}
\delta_\ell &= -(|\omega _M|+\ell)(1+|\omega _M|+\ell),\\
\Sigma &= |\omega_M|^2+|\omega_W|^2,\qquad
\Pi = |\omega_M|^2|\omega_W|^2.\\
\end{aligned}
\end{equation}
Specializing this expansion to the two cases shown in Fig.~\ref{fig:combined-winding}, we find
\begin{equation}\label{eq:FOZ_check_1}
\begin{aligned}
&\kappa_0(0,0;\mathfrak{t})=12\mathfrak{t}-6
-\frac{1}{\mathfrak{t}}-\frac{1}{2\mathfrak{t}^2}\\
&\quad-\frac{229}{720\mathfrak{t}^3}
-\frac{109}{480\mathfrak{t}^4}
-\frac{62999}{362880\mathfrak{t}^5}\\
&\quad-\frac{20159}{145152\mathfrak{t}^6}
-\frac{299803787}{2612736000\mathfrak{t}^7}\\
&\quad-\frac{72503387}{746496000\mathfrak{t}^8}
-\frac{173336436487}{2069286912000\mathfrak{t}^9}\\
&\quad-\frac{6744608267}{91968307200\mathfrak{t}^{10}}
+\mathcal{O}(\mathfrak{t}^{-11}),
\end{aligned}
\end{equation}
and
\begin{equation}
\begin{aligned}
&\kappa_0\br{0,1/2;\mathfrak{t}}=12\mathfrak{t}-6
-\frac{3}{4\mathfrak{t}}-\frac{3}{8\mathfrak{t}^2}\\
&\quad-\frac{61}{256\mathfrak{t}^3}
-\frac{87}{512\mathfrak{t}^4}
-\frac{1063}{8192\mathfrak{t}^5}\\
&\quad-\frac{1699}{16384\mathfrak{t}^6}
-\frac{4039069}{47185920\mathfrak{t}^7}\\
&\quad-\frac{6831883}{94371840\mathfrak{t}^8}
-\frac{94204909}{1509949440\mathfrak{t}^9}\\
&\quad-\frac{10989211}{201326592\mathfrak{t}^{10}}
+\mathcal{O}(\mathfrak{t}^{-11}).
\end{aligned}
\end{equation}
Using Eq.~\eqref{eq:rational_variables}, we can re-expand Eq.~\eqref{eq:FOZ_check_1} around $\eta=0$, extending the result of~\cite[Eq.~(5.20)]{Fateev:1992tk}:
\begin{equation}
\begin{aligned}
&\kappa_0\bigl(0,0;\mathfrak{t}(\eta)\bigr)
=\frac{3}{\eta}
-\frac{4}{45}\eta^3
+\frac{152}{2835}\eta^5\\
&\quad-\frac{16748}{637875}\eta^7
+\frac{150776}{12629925}\eta^9
+\mathcal{O}(\eta^{11}).
\end{aligned}
\end{equation}

\newpage
\paragraph{Small-$\mathfrak{q}$ expansions.}
We now turn to the small-$\mathfrak{q}$ expansion of the quantized accessory parameter. For $s\in\mathbb{Z}_{>1}$, we introduce
\begin{equation}
Z_s =
\sum_{\pm}\zeta\left(s,\frac{1+|\omega_M|\pm|\omega_W|}{2}\right)-2^s\zeta(s),\\
\end{equation}
where $\zeta(s,a)$ is the Hurwitz zeta function and $\zeta(s)=\zeta(s,1)$ is the Riemann zeta function. Recalling that $T = \frac{1}{4}\left(r_0 - \log \mathfrak{q}\right)$, with $r_0$ defined in~\eqref{eq:r0}, the first few terms of the $\mathfrak{q}$-expansion of $\kappa_\ell$ are
\begin{equation}
\label{eq:FOZ_check_2}
\kappa_{\ell}(\omega_M,\omega_W;\mathfrak{q}) = \kappa^{(0)}_\ell(T)+\kappa^{(1)}_\ell(T)\mathfrak{q}+\mathcal{O}(\mathfrak{q}^2),
\end{equation}
where, setting $\mathfrak{l}=\pi(\ell+1)$ and $\Omega=\Sigma^2-4\Pi$, we have
\begin{equation}
\begin{aligned}
&\kappa^{(0)}_\ell  =
6\left(|\omega_M|^2+|\omega_W|^2\right)
+\frac{3 \mathfrak{l}^2}{2T^2}+\frac{ \mathfrak{l}^4Z_3}{32T^5}\\
& \quad-\frac{3 \mathfrak{l}^6Z_5}{2560T^7}
+\frac{7 \mathfrak{l}^6Z_3^2}{6144T^8}
+\frac{3 \mathfrak{l}^8Z_7}{57344T^9}
\\
&\quad-\frac{9 \mathfrak{l}^8Z_3Z_5}{81920T^{10}}
+\mathcal{O}(T^{-11}),
\end{aligned}
\end{equation}
and
\begin{align}
&\kappa_\ell^{(1)} =
-3(1+2\Sigma+\Omega)
+\frac{3\mathfrak{l}^2(\Omega-1)}{4T^2}
+\frac{3\mathfrak{l}^2(\Omega+1)}{8T^3} \notag
\\
&
\quad-\frac{3\mathfrak{l}^4 \Omega}{16T^4}
+\frac{\mathfrak{l}^4}{64T^5}
 \bigl[(\Omega-1)Z_3-12\Omega\bigr] \notag
\\
&\quad+\frac{\mathfrak{l}^4}{T^6}
 \left[
 \frac{3\mathfrak{l}^2\Omega}{64}
 +\frac{5(\Omega+1)Z_3}{256}
 \right]
\\
&\quad+\frac{\mathfrak{l}^6}{T^7}
 \left[
 \frac{\Omega(9-Z_3)}{128}
 -\frac{3(\Omega-1)Z_5}{5120}
 \right] +\mathcal{O}(T^{-8}).\notag
\end{align}
For $\omega_M=\omega_W=\ell=0$, expressing $\mathfrak{q}$ and $T$ in terms of $\eta$, Eq.~\eqref{eq:FOZ_check_2} gives a double expansion in $(\eta+2\log2)^{-1}$ and $\re^{-4\eta}$, extending~\cite[Eq.~(5.23)]{Fateev:1992tk}.

\end{document}